\documentclass[aps,pra,showpacs,floatfix]{revtex4}
\usepackage{graphicx}
\usepackage{amsmath}
\usepackage{bm}

\begin{document}

\title{Excitation energies, oscillator strengths, and lifetimes of
levels along the gold isoelectronic sequence}

\author{ U.I. Safronova}
\email{usafrono@nd.edu}
\author{W. R. Johnson}
\email{johnson@nd.edu} \homepage{www.nd.edu/~johnson}
\affiliation{Department of Physics, 225 Nieuwland Science Hall\\
University of Notre Dame, Notre Dame, IN 46566}

\date{\today}
\begin{abstract}
Energies of $ns_{1/2}$ ($n$= 6-9), $np_j$ ($n$ = 6-8), $nd_j$
($n$= 6-7), and $5f_j$ states in neutral Au and Au-like ions with
nuclear charges $Z = 80 - 83$ are calculated using relativistic
many-body perturbation theory.  Reduced matrix elements,
oscillator strengths, transition rates and lifetimes are
determined for the 30 possible $nl_j-n'l'_{j'}$  electric-dipole
transitions. Results for  a limited number states $ns_{1/2}$,
$np_j$ ($n$= 6-7) and  $6d_j$ are obtained in the relativistic
single-double (SD) approximation, where single and double
excitations of Dirac-Fock wave functions are included to all
orders in perturbation theory. Using SD wave functions, accurate
values are obtained for energies of the eight lowest states and
for the fourteen possible electric-dipole matrix elements between
these states. With  the aid of the SD wave functions, we also
determine transition rates and oscillator strengths for the
fourteen transitions together with lifetimes of $6p_j$, $7p_j$,
and $6d_j$ levels. We investigate the hyperfine structure in Hg~II
and Tl~III. The hyperfine  $A$-values  are determined for
$6s_{1/2}$ and $6p_j$ states in $^{199}$Hg$^{+}$,
$^{201}$Hg$^{+}$, and $^{205}$Tl$^{2+}$ isotopes. These
calculations provide a theoretical benchmark for comparison with
experiment and theory.
 \pacs{31.15.Ar, 31.15.Md, 32.10.Fn, 32.70.Cs}
\end{abstract}
\maketitle
\section{Introduction}

This work continues earlier relativistic many-body
perturbation theory (MBPT) studies of energy levels of ions with
one valence electron outside a closed core.
In Refs.~\cite{li-en,na-en,cu-en,agmar} third-order MBPT was used to
calculate energies of the three lowest states ($ns_{1/2}$,
$np_{1/2}$, and $np_{3/2}$) in Li-, Na-, Cu-, and  Ag-like ions
along the respective isoelectronic sequences.
Third-order MBPT calculations of
$6s_{1/2}-6p_j$ transition amplitudes in Au-like ions up to $Z$=85
were performed in \cite{chou}. In the present paper,
we extend the previous calculations to obtain energies of
the sixteen lowest levels, $ns_{1/2}$ ($n$= 6-9), $np_j$ ($n$ =
6-8), $nd_j$ ($n$= 6-7), and $5f_j$ states, reduced matrix
elements, oscillator strengths, and transition rates for the 30
possible $nl_j-n'l'_{j'}$  electric-dipole transitions.
Additionally, we evaluate lifetimes of excited states for $6p_j$,
$7p_j$, and $6d_j$ levels
in neutral Au and Au-like ions with
nuclear charges $Z = 80 - 83$.

For the eight lowest states, $6s_{1/2}$, $6p_j$, $7p_j$, and
$6d_j$, we go beyond third-order MBPT and use the SD method, in
which single and double excitations of the Dirac-Fock (DF) wave
functions are summed to all orders, to evaluate energies,
transition rates, and lifetimes. The SD method was applied
previously to study properties of Li and Be$^+$
in \cite{blundell-li}, Na and Cs 
in \cite{Liu}, Cs
in \cite{blundell-cs}, Na-like ions with $Z$ ranging from 11 to 16
in \cite{safr-na}, and alkali-metal atoms Na, K, Rb, and Cs
in \cite{safr-alk}. It should be noted that the $n$ =1, 2, and 3
cores of Li-, Na-, and Cu-like ions are completely filled, by
contrast with Ag- and Au-like ions, where the $n=4$ and $n=5$
cores [Cu$^+$]$4s^24p^64d^{10}$ and [Nd]$5s^25p^65d^{10}$,
respectively, are incomplete.

Our results are compared with the theoretical results found in
Refs.~\cite{tl2,chou,hg1,au-like,au-tran} and
with experimental
measurements from Refs.~\cite{au-life,tl2-life,hg1-reader}.
Large-scale {\it ab initio} multiconfiguration Dirac-Fock (MCDF)
calculations were used in \cite{tl2} to obtain oscillator strengths
and hyperfine-structure parameters  for the lowest
few levels in Tl~III. Theoretical oscillator strengths and
hyperfine-structure parameters in Hg~II were determined in \cite{hg1}
using the MCDF method for resonance transitions.
A large number of transitions were also treated in \cite{hg1}
using a more
flexible, but less accurate, version of the MCDF method. These results
were used in stellar atmosphere models, assuming local
thermodynamic equilibrium (LTE) where a line-by-line investigation
was important. This was part of a program for studying chemically
peculiar stars using the Goddard High-Resolution Spectrograph
(GHRS) on board the Hubble Space Telescope. The ability
of an {\it ab initio} relativistic model potential approach with
explicit local exchange to produce oscillator strengths in
agreement with MCDF data was tested for seven transitions
in Au-like ions (Au~I, Hg~II, Tl~III, and Pb~IV) in \cite{au-like}.
In \cite{au-tran}, a
``weakest bound electron potential'' model was employed
to study transition probabilities in Au~I. In this
model, coupled equations were used to obtain parameters
$Z^*$, $n^*$, and $l^*$ used in the calculation of
transition probabilities.

Determinations of radiative lifetimes of excited states in neutral
gold using time-resolved vacuum-ultraviolet laser spectroscopy were
reported in Ref.~\cite{au-life}.
Recently, in Ref.~\cite{tl2-life}, measurements were reported for
lifetimes of the $6p_j$, $6pd_j$, $7s_{1/2}$, $7p_j$, $7d_j$,
and $5f_j$ levels in Tl~III  using beam-foil excitation.
As noted in \cite{tl2-life}, the Tl spectra in the
Hubble Space Telescope observations of peculiar stars required
oscillator-strength data for their interpretation. The spectrum
of Hg~II was observed from 500~\AA~ to 9880~\AA~with sliding
sparks and pulsed radio-frequency  discharges in \cite{hg1-reader},
where about 500 lines were classified
as transition between 114 energy levels. The observed
configurations ($5d^{10}nl$, $5d^96s5f$, $5d^96l6l'$, $5d^96l7l'$,
and $5d^86s^26p$) were theoretically interpreted by means of
HF calculations and least-squares fits of energy
parameters. The fitted parameters were then used
to calculate oscillator strengths for most of the classified
lines \cite{hg1-reader}.

In present paper, energies of $ns_{1/2}$ ($n$= 6-9), $np_j$ ($n$ =
6-8), $nd_j$ ($n$= 6-7), and $5f_j$ states in neutral Au and
Hg~II, Tl~III, Pb~IV, and Bi~V are obtained using relativistic
MBPT. Reduced matrix elements, oscillator
strengths, transition rates, and lifetimes are calculated for the
30 possible $nl_j-n'l'_{j'}$ electric-dipole transitions.
The eight lowest states
$ns_{1/2}$ ($n$= 6-7), $np_j$ ($n$= 6-7)
and  $6d_j$ are treated more accurately using the all-order SD method.
We also use the SD method to
obtain transition rates, oscillator strengths for electric dipole
transitions between these states, and lifetimes of the
$6p_j$, $7p_j$, and $6d_j$ levels.
Additionally, we evaluate hyperfine $A$-values for $6s_{1/2}$ and $6p_j$
states in $^{199}$Hg$^{+}$, $^{201}$Hg$^{+}$, and $^{205}$Tl$^{2+}$ isotopes.

\section{Third-order MBPT calculations of energies of Au-like ions}

Results of our third-order calculations of energies, which are
carried out following the pattern described in \cite{agmar}, are
summarized in Table~\ref{tab1}, where we list lowest-order,
Dirac-Fock energies $E^{(0)}$, first-order Breit energies
$B^{(1)}$, second-order Coulomb $E^{(2)}$ and Breit $B^{(2)}$
energies, third-order Coulomb energies  $E^{(3)}$, single-particle
Lamb shift corrections $E_\text{ LS}$, and the sum of the above
$E_{\text{tot}}$. The first-order Breit energies include
retardation, whereas the second-order
Breit energies are evaluated using the unretarded Breit operator. The
Lamb shift is approximated as the sum of the one-electron self
energy and the first-order vacuum-polarization energy. The
vacuum-polarization contribution is calculated from the Uehling
potential using the results of Fullerton and Rinker \cite{vacuum}.
The self-energy contribution is estimated for $s$, $p_{1/2}$ and
$p_{3/2}$ orbitals by interpolating among the values obtained by
\citet{mohr1,mohr2,mohr3} using Coulomb wave functions. For this
purpose, an effective nuclear charge $Z_\text{eff}$ is obtained by
finding the value of $Z_\text{eff}$ required to give a Coulomb
orbital with the same average $\langle r\rangle$ as the DHF
orbital.

We find that correlation corrections to energies in neutral Au and
 Au-like ions are large, especially for $6s$ states. For
example, $E^{(2)}$ is 28\% of $E^{(0)}$ and $E^{(3)}$ is 27\% of
$E^{(2)}$ for the $6s$ state of neutral Au. The ratio
$E^{(2)}/E^{(0)}$ decreases for the other (less penetrating) states
and for more highly charged ions but the second ratio
$E^{(3)}/E^{(2)}$ decreases only slowly for more excited states
 and even increases with $Z$ for the $6s$ state of
Bi~V.  Despite the slow convergence of the perturbation
expansion, the $6s$ energy from the present third-order MBPT
calculation is within 2.3\% of the measured ionization energy for
the $6s$ state of neutral Au and improves for higher valence
states and for more highly charged ions. 
The order of levels with respect to energy in Table~\ref{tab1} 
changes from ion to ion;
thus, for example, the $5f_{5/2}$ and $5f_{7/2}$ states are in the
twelfth and thirteenth places for neutral Au, in the tenth and
eleventh  places for Hg~II, and  in the seventh  and eighth
places for Bi~V. It should be mentioned that the difference in
energies of $5f_j$ and $7d_{j}$ states is less than 60 cm$^{-1}$
for Au~I, which may exceed the accuracy of the present calculations.

Below, we describe a few  numerical details of the calculation.
We use B-spline methods \cite{Bspline} to
generate a complete set of basis DHF wave functions for use in the
evaluation of MBPT expressions. For  Au~I and Bi~V, we use 40
splines of order $k=7$ for each angular momentum. The basis
orbitals are constrained to cavities of radii 85 a.u.\ and 55 a.u.\ for
Au~I and Bi~V, respectively. The cavity radius is scaled for
different ions; it is chosen large enough to accommodate all $6l_j$ and
$5f_j$ orbitals considered here and small enough that 40 splines
can approximate inner-shell DHF wave functions with good precision.
We use 35 out 40 basis orbitals for each partial wave in our
third-order energy calculations, since contributions from the five
highest-energy orbitals are negligible. The second-order
calculation includes partial waves up to $l_{\text{max}}=8$ and is
extrapolated to account for contributions from higher partial
waves.  A lower number of partial waves, $l_{\text{max}}=6$, is
used in the third-order calculation.  Since the asymptotic
$l$-dependence of the second- and third-order energies are similar
(both fall off as $l^{-4}$), we may use the second-order remainder
as a guide to extrapolating the third-order energy.

In Table~\ref{tab3}, we compare our results for energy levels
of the sixteen
single-particle states of interest in  Au~I, Hg~II, Tl~III, and Pb~IV
with recommended data from the
National Institute of Standards and Technology (NIST) database
\cite{nist}.
Although our results are generally in good
agreement with the NIST data, discrepancies were found. One cause
for these discrepancies is that fourth- and higher-order correlation
corrections are omitted in the theory. A second possible cause
is the omission of hole-particle-particle states in our single-particle 
model space.
The importance of $5d^{10}np$ +
$5d^{9}6s6p$ and $5d^{10}nf$ + $5d^{9}6s6p$ mixing for Au~I was
underlined by \citet{tl2-life}. Moreover,
\citet{hg1-reader} included hole-particle-particle states, $5d^96s5f$,
$5d^96l6l'$, $5d^96l7l'$ in calculations used to classify the observed 
spectrum of Hg~II.

\section{All-order SD calculations of energies of Au-like ions}

Results of our all-order SD calculations for the seven lowest states
of Au-like ions are presented in
Table~\ref{tab-esd}, where we list first-order (DHF) energies
$E^{(1)}$, SD correlation energies $E^\text{{SD}}$, omitted third-order terms
$E^{(3)}_\text{{extra}}$,
  first- and second-order Breit energies $B^{(n)}$, $n$ = 1, 2, single-particle
Lamb shift corrections $E_\text{ LS}$,
 totals $E_\text{ tot}$, and values from NIST, $E_\text{{NIST}}$
\cite{nist} for Au~I, Hg~II, Tl~III, Pb~IV, and Bi~V. The values
$E^\text{{SD}}$ are evaluated using the SD method.
The SD equations are set up in a finite basis and solved
iteratively to give the single- and double-excitation coefficients
and the correlation energy $E^\text{{SD}}$.
The contributions  $E^{(3)}_\text{{extra}}$
 in Table~\ref{tab-esd} accounts for that part of the
third-order MBPT correction not included in the SD
energy (see Eq.~(2.17) in Ref.~\cite{Safronova}).
The basis orbitals used to define
the single-particle SD states are linear combinations of B-splines.

As can be seen by comparing  Tables~\ref{tab1} and
\ref{tab-esd}, the dominant contribution to  $E^\text{{SD}}$ is
the second-order term $E^{(2)}$. We have already
mentioned the importance of higher partial wave contributions to
$E^{(2)}$ and
$E^{(3)}$. In the SD calculation, we include partial
waves through $l$=6 in $E^\text{{SD}}$ and use
the difference $E^{(2)}_{l\leq 8}$ - $E^{(2)}_{l\leq 6}$ to
estimate the accuracy of our result.

The columns with headings $\delta E$ in Tables~\ref{tab3} and
\ref{tab-esd} show the differences between our {\it ab initio}
 results and the recommended NIST data
\cite{nist}. As can be seen, the SD
results agree better with the recommended NIST data \cite{nist}
than do the third-order MBPT results, especially for ionization
potential. This confirms our
previous comments concerning the slow convergence of perturbation
theory and illustrates the importance of
fourth- and higher-order correlation corrections.
Those differences between the present theoretical
results and the recommended NIST data \cite{nist}
that were not improved by the SD method are most probably due
to the omission of hole-particle-particle states
mentioned previously.

\section{Dipole matrix elements, oscillator strengths,
and lifetimes in Au-like ions}

Transition matrix elements provide another test of the quality of
atomic-structure calculations and another measure of the size of
correlation corrections.  Reduced matrix elements of the dipole
operator in first-, second-, third-, and all-order perturbation
theory between low-lying states of Au~I and Bi~V are presented
in Table~\ref{tab-dip}. The first-order reduced matrix elements
$Z^{(1)}$ are obtained from length-form DHF calculations.
Length-form and velocity-form matrix elements differ typically by
1 - 3\%. Second-order matrix element in the table, $Z^{(2)}$,
which include $Z^{(1)}$, are extended to include the second-order
correction associated with the random-phase approximation (RPA).

The third-order matrix elements $Z^{(3)}$ include $Z^\text{ (RPA)}$
(all higher-order RPA corrections), 
Bruekner-orbital $Z^\text{ (BO)}$, structural radiation $Z^\text{
(SR)}$, and normalization $Z^\text{ (NORM)}$  corrections,
described, for example in Refs.~\cite{dip3,Safronova}. As can be seen in
Table~\ref{tab-dip}, RPA corrections are very large, 10-40\%,
being smallest for the $6p_{j}$-$7s_{1/2}$ transitions. 

Electric-dipole matrix elements evaluated in the SD approximation
are given in  columns headed $Z^\text{{(SD)}}$  in
Table~\ref{tab-dip}. A detailed discussion of calculations of matrix elements
in the SD approximation is found in Ref.~\cite{blundell-li}. 
It should be noted
that SD matrix elements $Z^\text{{SD}}$ include $Z^{(3)}$ completely,
along with important fourth- and higher-order corrections. Those
fourth-order corrections omitted from SD matrix
elements were discussed recently by \citet{der-4}.

We carried out third-order calculations for a much longer list
of transitions than presented in Table~\ref{tab-dip}. The results
of our third-order calculations are
summarized in Table~\ref{tab-osc}, where we list oscillator
strengths for thirty $6s-6p$, $6s-7p$, $6s-8p$, $6p-6d$, $6p-7s$,
$6p-8s$, $6d-7p$, $6d-5f$, and $5f-7d$ transitions in  Au~I,
Hg~II, Tl~III, Pb~IV,  and Bi~V. For each ion the oscillator
strengths are calculated in first- and third-order MBPT;
$f^{(1)}$ and $f^{(3)}$. It should be noted, that we use
theoretical energies in the same order of approximation (columns
$E^{(0)}$ and $E_\text{tot}$ from Table~\ref{tab1}) to calculate
$f^{(1)}$ and $f^{(3)}$, respectively. In Table~\ref{tab-osc}, we
have marked $6d-7p$ transition for Au~I with an asterisk,
since the order of initial and final states is reversed
for these transitions.
 As can be seen from
Table~\ref{tab-osc}, the difference between $f^{(1)}$ and
$f^{(3)}$ ranges from 10\% to 40\% for cases with large values of
oscillator strengths ($6s-6p$, $7p-8s$, $6p-6d$, and $6d-5f$
transitions). For cases with small oscillator
strengths($6s-7p$ and $5f-7d$ transitions), 
differences between $f^{(1)}$ and $f^{(3)}$ can be a factor of 10.
In the case of $6s-7p$ transitions, 
large differences between dipole matrix element
$Z^{(3)}$ and $Z^{(1)}$ are responsible for the
large differences between $f^{(3)}$ and $f^{(1)}$ (see Table~\ref{tab-dip}),
whereas, in the case of $5f-7d$ transitions, large differences in transition energies
(see Table~\ref{tab1}) are responsible for the large $f^{(3)}$ -- $f^{(1)}$ differences.

In  Table~\ref{tab1-comp}, we compare our oscillator strength data
with theoretical results from
Ref.~\cite{au-like} and with available experimental measurements.
The oscillator strengths $f^\text{SD}$ in the table are calculated in
the SD approximation. It should be noted, that theoretical SD
energies  $E_\text{tot}$ from Table~\ref{tab-esd} are used to calculate
$f^\text{SD}$. Theoretical oscillator strengths $f^\text{theor}$ from
~\cite{au-like} were calculated using polarizable frozen
ion-like Dirac-Fock method (DF+CP). The experimental data
$f^\text{expt}$  listed in Table~\ref{tab1-comp} are from
Ref.~\cite{au-like} and references therein. The accuracy of
the experimental $f$-values is not very high and in some cases the difference
between results from different measurements  is larger than the
difference between experimental and theoretical results.

We calculate lifetimes of $6p_j$, $6d_j$, $7s_{1/2}$, and  $7p_j$
levels in neutral Au and in Au-like ions with $Z$ = 80--83  using
both third-order MBPT and SD results for dipole matrix
elements and energies.
We list lifetimes, $\tau^\text{(SD)}$ and
wavelengths, $\lambda^\text{(SD)}$, obtained by SD method in
Table~\ref{tab-life}.  In this table,
 we compare our lifetime data with
available experimental measurements that are primarily obtained
for $6p_j$ levels. The experimental data are from
Ref.~\cite{tl2-life} and references therein.  Even for these
lowest-lying states we have 5 - 10\% disagreement between our
$\tau^\text{(SD)}$ and $\tau^\text{expt}$ for both, $6p_{1/2}$ and
$6p_{3/2}$ states in Au~I, Hg~II, and Tl~III, and for the $6p_{3/2}$
state in Bi~V. This is somewhat strange since we have perfect
agreement between $\lambda^\text{(SD)}$ and $\lambda^\text{expt}$
for all of these levels, as can be seen in Table~\ref{tab-life}.

\section{Hyperfine constants for Au-like ions}

Calculations of hyperfine constants follow the same pattern as
the calculations of reduced dipole matrix elements described in
the previous section.
The magnetic moments and nuclear spins used in present
calculations are taken from \cite{web}.  In Table~\ref{tab-hyp},
we give the magnetic-dipole hyperfine constant $A$ for
$^{205}$Tl~III, $^{199}$Hg~II, and $^{201}$Hg~II and compare with
available theoretical and experimental data from Refs.~\cite{tl2,hg1},
 and references therein. In this table, we present
 the first-order $A^\text{(DF)}$, third-order $A^{(3)}$,
and all-order $A^\text{(SD)}$  values for $6s_{1/2}$,
$6p_{1/2}$, and  $6p_{3/2}$ levels. As discussed before for
dipole matrix elements,  third-order hyperfine
constants  $A^{(3)}$  include RPA $A^\text{(RPA)}$, Bruekner-orbital 
$A^\text{(BO)}$, structural radiation $A^\text{ (SR)}$, 
and normalization
$A^\text{ (NORM)}$ corrections.  The
difference between third- and first-order contributions 
as can be seen from Table~\ref{tab-hyp}, is 20 - 30~\%.
The Bruekner-orbital term $A^\text{(BO)}$
gives the largest correction in all cases. 
 As can be seen from Table~\ref{tab-hyp}, the all-order SD
$A^\text{(SD)}$ results are in better agreement with 
the theoretical results
$A^\text{(theor)}$ from Refs.~\cite{tl2,hg1}, than are the
third-order values
$A^{(3)}$.  Moreover, 
$A^\text{(SD)}$ and $A^\text{(theor)}$
agree better with each other than with
experimental data $A^\text{(expt)}$.

\section{Conclusion}
In summary,  a systematic  MBPT study of the energies of
$ns_{1/2}$ ($n$= 6-9), $np_j$ ($n$ = 6-8), $nd_j$ ($n$= 6-7), and
$5f_j$ states in neutral Au and Au-like ions with nuclear charges
$Z = 80 - 83$  is presented. The energy calculations are
in good agreement with existing experimental energy data and
provide a  theoretical reference database for the line
identification. A systematic relativistic MBPT study of reduced
matrix elements and oscillator strengths for the 30 possible
$6s-6p$, $6s-7p$, $6s-8p$, $6p-6d$, $6p-7s$, $6p-8s$, $6d-7p$,
$6d-5f$, and $5f-7d$ transitions in  Au~I, Hg~II, Tl~III, Pb~IV,
and Bi~V  is conducted. Both length and velocity forms of matrix
elements are evaluated. Small differences between length and
velocity-form calculations, caused by the nonlocality of the DF
potential, are found in second order. However, including
third-order corrections with full RPA leads to complete agreement
between the length- and velocity-form results.  Hyperfine
$A$-values are presented for $6s_{1/2}$ and $6p_j$ states in
$^{199}$Hg$^{+}$, $^{201}$Hg$^{+}$, and $^{205}$Tl$^{2+}$ ions.

We believe that our energies and transition rates will be useful
in analyzing existing experimental data and in planning new
experiments. There remains a paucity of experimental data for many
of the higher ionized members of this sequence, both for term
energies and for  lifetimes.

\begin{acknowledgments}
The work of W.R.J.  was supported in part by National Science
Foundation Grant No.\ PHY-01-39928.
\end{acknowledgments}

\begin{table*}
\caption{\label{tab1} Contributions to energy levels of
Au-like ions  in cm$^{-1}$.}
\begin{ruledtabular}
\begin{tabular}{lrrrrrrrcrrrrrrr}
\multicolumn{1}{c}{$nlj$ } &
\multicolumn{1}{c}{$E^{(0)}$} &
\multicolumn{1}{c}{$E^{(2)}$} &
\multicolumn{1}{c}{$E^{(3)}$} &
\multicolumn{1}{c}{$B^{(1)}$} &
\multicolumn{1}{c}{$B^{(2)}$} &
\multicolumn{1}{c}{$E_\text{ LS}$}&
\multicolumn{1}{c}{$E_\text{ tot}$} &
\multicolumn{1}{c}{} &
\multicolumn{1}{c}{$E^{(0)}$} &
\multicolumn{1}{c}{$E^{(2)}$} &
\multicolumn{1}{c}{$E^{(3)}$} &
\multicolumn{1}{c}{$B^{(1)}$} &
\multicolumn{1}{c}{$B^{(2)}$} &
\multicolumn{1}{c}{$E_\text{ LS}$}&
\multicolumn{1}{c}{$E_\text{ tot}$} \\
\hline
\multicolumn{8}{c}{Au~I } &
\multicolumn{1}{c}{       } &
\multicolumn{7}{c}{Hg~II }\\
$6s_{1/2}$&  -60270& -16611&  4493&  253& -595&  42&  -72686&&  -136471& -18393&  5382&  389&  -756&  79& -149770\\
$6p_{1/2}$&  -29362&  -6889&  1431&  105& -187&  -1&  -34903&&   -89694& -11013&  2670&  269&  -375&  -2&  -98144\\
$6p_{3/2}$&  -26691&  -5187&  1029&   54& -123&   0&  -30919&&   -82029&  -8885&  2096&  153&  -272&   1&  -88937\\
$7s_{1/2}$&  -18265&  -1918&   525&   31&  -66&   2&  -19691&&   -52828&  -3314&   974&   74&  -129&   7&  -55216\\
$7p_{1/2}$&  -12388&  -1353&   279&   24&  -41&   0&  -13479&&   -40892&  -2478&   594&   70&   -90&   0&  -42796\\
$7p_{3/2}$&  -11708&  -1153&   223&   14&  -31&   0&  -12654&&   -38732&  -2207&   486&   44&   -74&   0&  -40483\\
$5f_{5/2}$&   -6861&    -55&    11&    0&    0&   0&   -6905&&   -27610&   -571&   108&    0&    -3&   0&  -28076\\
$5f_{7/2}$&   -6862&    -55&    11&    0&    0&   0&   -6906&&   -27625&   -571&   105&    0&    -3&   0&  -28094\\
$6d_{3/2}$&  -11930&   -452&    75&    3&   -8&   0&  -12311&&   -44308&  -1960&   362&   28&   -58&   0&  -45935\\
$6d_{5/2}$&  -11875&   -429&    68&    3&   -9&   0&  -12243&&   -43866&  -1845&   322&   23&   -60&   0&  -45426\\
$7d_{3/2}$&   -6714&   -190&    30&    2&   -4&   0&   -6877&&   -25173&   -770&   128&   13&   -27&   0&  -25829\\
$7d_{5/2}$&   -6689&   -180&    26&    1&   -4&   0&   -6845&&   -24970&   -733&    96&   11&   -28&   0&  -25624\\
$8s_{1/2}$&   -9126&   -626&   162&   11&  -22&   0&   -9602&&   -28839&  -1226&   345&   29&   -50&   2&  -29740\\
$8p_{1/2}$&   -6899&   -513&    93&   10&  -15&   0&   -7324&&   -23794&  -1009&   219&   30&   -35&   0&  -24589\\
$8p_{3/2}$&   -6618&   -452&    76&    6&  -12&   0&   -7000&&   -22844&   -923&   184&   20&   -29&   0&  -23591\\
$9s_{1/2}$&   -5483&   -286&    74&    5&  -11&   0&   -5700&&   -18214&   -601&   169&   14&   -24&   1&  -18655\\
\multicolumn{8}{c}{Tl~III } &
\multicolumn{1}{c}{       } &
\multicolumn{7}{c}{Pb~VI }\\
$6s_{1/2}$& -225476& -19585&  6142&  529& -899& 120& -239170&&  -325644& -20578&  6288&  678&-1036& 167& -340126\\
$6p_{1/2}$& -165181& -13419&  3656&  443& -529&  -3& -175033&&  -252787& -15218&  4132&  631& -671&  -4& -263918\\
$6p_{3/2}$& -151926& -11117&  2622&  261& -399&   1& -160558&&  -233342& -12781&  3184&  379& -518&   2& -243077\\
$7s_{1/2}$&  -98041&  -4408&  1373&  125& -191&  16& -101127&&  -152137&  -5336&  1627&  183& -253&  27& -155888\\
$7p_{1/2}$&  -80050&  -3484&   833&  130& -143&  -1&  -82715&&  -128093&  -4406&  1191&  201& -198&  -1& -131306\\
$7p_{3/2}$&  -75917&  -3135&   711&   84& -121&   0&  -78378&&  -121553&  -3988&   960&  131& -170&   0& -124619\\
$5f_{5/2}$&  -63404&  -2143&   350&    5&  -31&   0&  -65223&&  -116996&  -5061&   734&   31& -152&   0& -121445\\
$5f_{7/2}$&  -63501&  -2130&   334&    4&  -31&   0&  -65325&&  -117248&  -4952&   686&   24& -151&   0& -121642\\
$6d_{3/2}$&  -92016&  -3631&   725&   74& -135&   0&  -94982&&  -152198&  -5207&  1103&  135& -224&   0& -156390\\
$6d_{5/2}$&  -90884&  -3438&   637&   59& -136&   0&  -93762&&  -150159&  -4960&   964&  107& -223&   0& -154271\\
$7d_{3/2}$&  -53047&  -1451&   190&   34&  -61&   0&  -54335&&   -89044&  -2221&   300&   62& -101&   0&  -91004\\
$7d_{5/2}$&  -52529&  -1454&   120&   27&  -62&   0&  -53898&&   -88113&  -1837&   220&   49& -102&   0&  -89783\\
$8s_{1/2}$&  -56158&  -1763&   511&   53&  -79&   4&  -57431&&   -89982&  -2261&   663&   81& -111&   8&  -91600\\
$8p_{1/2}$&  -48053&  -1481&   349&   59&  -59&   0&  -49186&&   -78669&  -1982&   487&   95&  -88&   0&  -80157\\
$8p_{3/2}$&  -46146&  -1380&   292&   39&  -50&   0&  -47245&&   -75543&  -1836&   386&   63&  -75&   0&  -77004\\
$9s_{1/2}$&  -36493&   -891&   258&   27&  -40&   2&  -37137&&   -59627&  -1241&   361&   43&  -59&   4&  -60520\\
\multicolumn{8}{c}{Bi~V } &
\multicolumn{1}{c}{       } &
\multicolumn{7}{c}{Bi~V  }\\
$6s_{1/2}$& -436067& -21459&  6581&  836&-1171& 222& -451059&$6d_{3/2}$& -223271&  -6665&  1483&  209& -318&   0& -228562\\
$6p_{1/2}$& -351174& -16712&  4641&  832& -808&  -5& -363227&$6d_{5/2}$& -220145&  -6375&  1304&  164& -317&   0& -225368\\
$6p_{3/2}$& -324912& -14154&  3907&  506& -633&   3& -335283&$7d_{3/2}$& -132398&  -2003&   350&   96& -146&   0& -134098\\
$7s_{1/2}$& -214192&  -6158&  1882&  248& -315&  43& -218492&$7d_{5/2}$& -130963&  -2590&   250&   76& -145&   0& -133371\\
$7p_{1/2}$& -184093&  -5267&  1464&  283& -255&  -1& -187869&$8s_{1/2}$& -129701&  -2715&   798&  114& -143&  14& -131633\\
$7p_{3/2}$& -174729&  -4790&  1203&  186& -220&   0& -178351&$8p_{1/2}$& -115073&  -2481&   620&  137& -116&  -1& -116913\\
$5f_{5/2}$& -190874&  -8480&  1202&  101& -416&   0& -198466&$8p_{3/2}$& -110475&  -2322&   420&   92&  -97&   0& -112382\\
$5f_{7/2}$& -191165&  -8165&  1109&   77& -407&   0& -198552&$9s_{1/2}$&  -87216&  -1528&   441&   63&  -79&   6&  -88313\\
\end{tabular}
\end{ruledtabular}
\end{table*}

\begin{table*}
\caption{\label{tab3} Comparison of the energies of the $nl_j$  states in Au-like ions  with experimental data  \protect\cite{nist}.
 Units: cm$^{-1}$.}
\begin{ruledtabular}
\begin{tabular}{lrrrrrrrrrrrrrr}
\multicolumn{1}{c}{$nlj$ } &
\multicolumn{1}{c}{$E_{\text{tot}}$} &
\multicolumn{1}{c}{$E_{\text{expt}}$} &
\multicolumn{1}{c}{$\delta E$} &
\multicolumn{1}{c}{$E_{\text{tot}}$} &
\multicolumn{1}{c}{$E_{\text{expt}}$} &
\multicolumn{1}{c}{$\delta E$} &
\multicolumn{1}{c}{$E_{\text{tot}}$} &
\multicolumn{1}{c}{$E_{\text{expt}}$} &
\multicolumn{1}{c}{$\delta E$} &
\multicolumn{1}{c}{$E_{\text{tot}}$} &
\multicolumn{1}{c}{$E_{\text{expt}}$} &
\multicolumn{1}{c}{$\delta E$}\\
\hline
\multicolumn{4}{c}{Au~I } &
\multicolumn{3}{c}{Hg~II } &
\multicolumn{3}{c}{Tl~III } &
\multicolumn{3}{c}{Pb~IV } \\
$6s_{1/2}$& -72686& -74410& 1724&-149770&-151280&   1510&-239170&-240600&   1430&-340126& -341350&  1224\\
$6p_{1/2}$& -34903& -37056& 2153& -98144& -99795&   1651&-175033&-176443&   1410&-263918& -265192&  1274\\
$6p_{3/2}$& -30919& -33236& 2317& -88937& -90672&   1735&-160558&-161630&   1072&-243077& -244131&  1054\\
$7s_{1/2}$& -19691& -19925&  234& -55216& -55566&    350&-101127&-101391&    264&-155888& -156247&   359\\
$7p_{1/2}$& -13479& -14377&  898& -42796& -42982&    186& -82715& -82748&     33&-131306&        &      \\
$7p_{3/2}$& -12654& -13681& 1027& -40483& -39310&  -1173& -78378& -77066&  -1312&-124619&        &      \\
$5f_{5/2}$&  -6905&  -6925&   20& -28076& -27871&   -205& -65223& -63645&  -1578&-121445&        &      \\
$5f_{7/2}$&  -6906&  -6920&   14& -28094& -28128&     34& -65325& -65007&   -318&-121642&        &      \\
$6d_{3/2}$& -12311& -12458&  147& -45935& -46297&    362& -94982& -95245&    263&-156390&  -156791&  401\\
$6d_{5/2}$& -12243& -12376&  133& -45426& -45737&    311& -93762& -93931&    169&-154271&  -154533&  262\\
$7d_{3/2}$&  -6877&  -6941&   64& -25829& -25956&    127& -54335& -54244&    -91& -91004&   -90948&  -56\\
$7d_{5/2}$&  -6845&  -6899&   54& -25624& -25702&     78& -53898& -53652&   -246& -89783&   -89931&  148\\
$8s_{1/2}$&  -9602&  -9668&   66& -29740& -29864&    124& -57431& -57413&    -18& -91600&   -91716&  116\\
$8p_{1/2}$&  -7324&  -7805&  481& -24589& -24338&   -251& -49186&       &       & -80157&        &      \\
$8p_{3/2}$&  -7000&  -7500&  500& -23591& -23482&   -109& -47245&       &       & -77004&        &      \\
$9s_{1/2}$&  -5700&  -5729&   29& -18655& -18721&     66& -37137&       &       & -60520&        &      \\
\end{tabular}
\end{ruledtabular}
\end{table*}

\begin{table*}
\caption{\label{tab-esd} First-order (DHF) energies $E^{(1)}$,
single-double Coulomb energies $E^\text{{SD}}$,
$E^{(3)}_\text{{extra}}$,
  first- and second-order Breit, Lamb correction $E_\text{ LS}$,
and totals $E_\text{ tot}$ for Au~I, Hg~II, Tl~III, Pb~IV, and Bi~V are compared with
 energies $E_\text{{NIST}}$ \protect\cite{nist}, ($E_\text{ tot}$
 -$E_\text{{NIST}}$ =$\delta E$). Units,
cm$^{-1}$.}
\begin{ruledtabular}
\begin{tabular}{lrrrrrrrrr}
\multicolumn{1}{c}{$nlj$ } &
\multicolumn{1}{c}{$E^{(1)}$} &
\multicolumn{1}{c}{$E^\text{{SD}}$} &
\multicolumn{1}{c}{$E^{(3)}_\text{{extra}}$} &
\multicolumn{1}{c}{$B^{(1)}$} &
\multicolumn{1}{c}{$B^{(2)}$} &
\multicolumn{1}{c}{$E_\text{ LS}$}&
\multicolumn{1}{c}{$E_\text{ tot}$} &
\multicolumn{1}{c}{$E_\text{{NIST}}$} &
\multicolumn{1}{c}{$\delta E$} \\
\hline
\multicolumn{9}{c}{Au~I}\\
$6s_{1/2}$& -60270& -14885&   1243&    253&    -595&   42&  -74210& -74410&    200\\
$6p_{1/2}$& -29362&  -7891&    453&    105&    -187&   -1&  -36883& -37056&    173\\
$6p_{3/2}$& -26691&  -6393&    367&     54&    -123&    0&  -32786& -33236&    450\\
$7s_{1/2}$& -18265&  -1705&    173&     31&     -66&    2&  -19831& -19925&     94\\
$7p_{1/2}$& -12388&  -1294&     88&     24&     -41&    0&  -13611& -14377&    766\\
$7p_{3/2}$& -11708&  -1202&     78&     14&     -31&    0&  -12848& -13681&    833\\
$6d_{3/2}$& -11930&   -541&     45&      3&      -8&    0&  -12431& -12458&     27\\
$6d_{5/2}$& -11875&   -535&     41&      3&      -9&    0&  -12376& -12376&      0\\
\multicolumn{9}{c}{Hg~II}\\
$6s_{1/2}$&-136471& -15631&   1594&    389&    -756&   79& -150796&-151280&    484\\
$6p_{1/2}$& -89694& -10441&    840&    269&    -375&   -2&  -99402& -99795&    393\\
$6p_{3/2}$& -82029&  -8780&    712&    153&    -272&    1&  -90216& -90672&    456\\
$7s_{1/2}$& -52828&  -2870&    333&     74&    -129&    7&  -55414& -55566&    152\\
$7p_{1/2}$& -40892&  -2301&    203&     70&     -90&    0&  -43011& -42982&    -29\\
$7p_{3/2}$& -38732&  -2201&    182&     44&     -74&    0&  -40781& -39310&  -1471\\
$6d_{3/2}$& -44308&  -2067&    172&     28&     -58&    0&  -46233& -46297&     64\\
$6d_{5/2}$& -43866&  -2064&    157&     23&     -60&    0&  -45810& -45737&    -73\\
\multicolumn{9}{c}{Tl~III}\\
$6s_{1/2}$&-225476& -16312&   1887&    529&    -899&  120& -240151&-240600&    449\\
$6p_{1/2}$&-165181& -11980&   1159&    443&    -529&   -3& -176091&-176443&    352\\
$6p_{3/2}$&-151926& -10204&    994&    261&    -399&    1& -161273&-161630&    357\\
$7s_{1/2}$& -98041&  -3753&    472&    125&    -191&   16& -101372&-101391&     19\\
$7p_{1/2}$& -80050&  -3156&    297&    130&    -143&   -1&  -82921& -82748&   -173\\
$7p_{3/2}$& -75917&  -3055&    269&     84&    -121&    0&  -78739& -77066&  -1673\\
$6d_{3/2}$& -92016&  -3688&    315&     74&    -135&    0&  -95450& -95245&   -205\\
$6d_{5/2}$& -90884&  -3931&    294&     59&    -136&    0&  -94598& -93931&   -667\\
\multicolumn{9}{c}{Pb~IV}\\
$6s_{1/2}$&-325644& -16977&   1985&    678&   -1036&  167& -340827&-341350&    523\\
$6p_{1/2}$&-252787& -13201&   1337&    631&    -671&   -4& -264695&-265192&    497\\
$6p_{3/2}$&-233342& -11315&   1161&    379&    -518&    2& -243634&-244131&    497\\
$7s_{1/2}$&-152137&  -4495&    565&    183&    -253&   27& -156109&-156247&    138\\
$7p_{1/2}$&-128093&  -3910&    421&    201&    -198&   -1& -131580&-341350&       \\
$7p_{3/2}$&-121553&  -3825&    383&    131&    -170&    0& -125034&-341350&       \\
$6d_{3/2}$&-152198&  -5128&    458&    135&    -224&    0& -156956&-156791&   -165\\
$6d_{5/2}$&-150159&  -5654&    433&    107&    -223&    0& -155496&-154533&   -963\\
\multicolumn{9}{c}{Bi~V}\\
$6s_{1/2}$&-436067& -17627&   2117&    836&   -1171&  222& -451690&-451700&     10\\
$6p_{1/2}$&-351174& -14264&   1520&    832&    -808&   -5& -363899&-363946&     47\\
$6p_{3/2}$&-324912& -12272&   1327&    506&    -633&    3& -335981&-336026&     45\\
$7s_{1/2}$&-214192&  -5155&    656&    248&    -315&   43& -218715&-218360&   -355\\
$7p_{1/2}$&-184093&  -4527&    520&    283&    -255&   -1& -188074&-181075&  -6999\\
$7p_{3/2}$&-174729&  -4554&    476&    186&    -220&    0& -178842&-170619&  -8223\\
$6d_{3/2}$&-223271&  -6378&    596&    209&    -318&    0& -229162&-222411&  -6751\\
$6d_{5/2}$&-220145&  -7019&    568&    164&    -317&    0& -226748&-219160&  -7588\\
\end{tabular}
\end{ruledtabular}
\end{table*}

\begin{table*}
\caption{\label{tab-dip} Reduced matrix elements of the dipole
operator in first-, second-, third-, and all-order perturbation
theory in  Au~I and Bi~V. }
\begin{ruledtabular}
\begin{tabular}{llrrrrrrrr}
\multicolumn{2}{c}{Transition}&
\multicolumn{1}{c}{$Z^{(1)}$ }&
\multicolumn{1}{c}{$Z^{(2)}$ }&
\multicolumn{1}{c}{$Z^{(3)}$ }&
\multicolumn{1}{c}{$Z^\text{{(SD)}}$ }&
\multicolumn{1}{c}{$Z^{(1)}$ }&
\multicolumn{1}{c}{$Z^{(2)}$ }&
\multicolumn{1}{c}{$Z^{(3)}$ }&
\multicolumn{1}{c}{$Z^\text{{(SD)}}$ }\\
\hline
\multicolumn{5}{c}{Au~I}&
\multicolumn{4}{c}{Bi~V}\\
$6s_{1/2}$&$ 6p_{1/2}$&  2.713&   2.114&   1.829&   1.822&  1.699&  1.242&  1.290&  1.320\\
$6s_{1/2}$&$ 6p_{3/2}$&  3.704&   2.972&   2.543&   2.546&  2.383&  1.784&  1.840&  1.877\\
$6p_{1/2}$&$ 6d_{3/2}$&  4.792&   4.556&   3.080&   3.363&  2.551&  2.124&  2.106&  2.131\\
$6p_{3/2}$&$ 6d_{3/2}$&  2.557&   2.447&   1.823&   1.859&  1.292&  1.092&  1.086&  1.095\\
$6p_{3/2}$&$ 6d_{5/2}$&  7.607&   7.286&   5.385&   5.545&  3.833&  3.253&  3.229&  2.918\\
$6p_{1/2}$&$ 7s_{1/2}$&  3.119&   3.113&   2.257&   2.317&  0.881&  0.920&  0.867&  0.867\\
$6p_{3/2}$&$ 7s_{1/2}$&  5.417&   5.355&   4.247&   4.120&  1.697&  1.692&  1.619&  1.619\\
$6s_{1/2}$&$ 7p_{1/2}$&  0.398&   0.163&   0.035&   0.017&  0.080&  0.260&  0.256&  0.240\\
$6s_{1/2}$&$ 7p_{3/2}$&  0.799&   0.479&   0.299&   0.252&  0.121&  0.163&  0.190&  0.168\\
$7s_{1/2}$&$ 7p_{1/2}$&  7.256&   7.137&   6.739&   6.793&  3.253&  3.058&  3.015&  3.030\\
$6d_{3/2}$&$ 7p_{1/2}$& 12.337&  12.275&  11.973&  11.657&  2.976&  2.912&  2.846&  2.769\\
$7s_{1/2}$&$ 7p_{3/2}$&  9.596&   9.467&   8.943&   9.105&  4.458&  4.219&  4.160&  4.094\\
$6d_{3/2}$&$ 7p_{3/2}$&  5.404&   5.388&   5.325&   5.175&  1.174&  1.171&  1.139&  1.069\\
$6d_{5/2}$&$ 7p_{3/2}$& 16.371&  16.319&  16.161&  15.496&  3.705&  3.682&  3.590&  2.745\\
 \end{tabular}
\end{ruledtabular}
\end{table*}

\begin{table*}
\caption{\label{tab-osc} Oscillator strengths ($f$) for
transitions in Au-like ions  calculated in lowest order (DF
approximation) $f^{(1)}$ and third order $f^{(3)}$ MBPT. }
\begin{ruledtabular}
\begin{tabular}{llllllllllll}
\multicolumn{2}{c}{} &
\multicolumn{2}{c}{Au~I } &
\multicolumn{2}{c}{Hg~II} &
\multicolumn{2}{c}{Tl~III} &
\multicolumn{2}{c}{Pb~IV} &
\multicolumn{2}{c}{Bi~V} \\
\multicolumn{2}{c}{$nl_{j}- n'l'_{j'}$} &
\multicolumn{1}{c}{$f^{(1)}$} & \multicolumn{1}{c}{$f^{(3)}$} &
\multicolumn{1}{c}{$f^{(1)}$} & \multicolumn{1}{c}{$f^{(3)}$} &
\multicolumn{1}{c}{$f^{(1)}$} & \multicolumn{1}{c}{$f^{(3)}$} &
\multicolumn{1}{c}{$f^{(1)}$} & \multicolumn{1}{c}{$f^{(3)}$} &
\multicolumn{1}{c}{$f^{(1)}$} &
\multicolumn{1}{c}{$f^{(3)}$} \\
\hline
$6s_{1/2}$&$ 6p_{1/2}$&  0.3454  &0.1919  &0.3717 &0.2071 &0.3754 &0.2145 &0.3747 &0.2194 &0.3723 &0.2221\\
$6s_{1/2}$&$ 6p_{3/2}$&  0.6996  &0.4103  &0.8372 &0.4927 &0.8950 &0.5345 &0.9316 &0.5687 &0.9587 &0.5956\\
$6p_{1/2}$&$ 6d_{3/2}$&  0.6080  &0.3255  &0.8942 &0.6402 &1.0722 &0.7809 &1.1864 &0.8588 &1.2646 &0.9074\\
$6p_{3/2}$&$ 6d_{3/2}$&  0.0733  &0.0469  &0.1009 &0.0766 &0.1158 &0.0883 &0.1241 &0.0934 &0.1289 &0.0956\\
$6p_{3/2}$&$ 6d_{5/2}$&  0.6510  &0.4113  &0.8948 &0.6765 &1.0347 &0.7896 &1.1174 &0.8432 &1.1686 &0.8702\\
$6p_{1/2}$&$ 7s_{1/2}$&  0.1640  &0.1177  &0.1627 &0.1458 &0.1620 &0.1558 &0.1618 &0.1616 &0.1616 &0.1652\\
$6p_{3/2}$&$ 7s_{1/2}$&  0.1878  &0.1538  &0.2130 &0.1926 &0.2252 &0.2116 &0.2345 &0.2236 &0.2422 &0.2324\\
$6s_{1/2}$&$ 7p_{1/2}$&  0.0115  &0.0001  &0.0013 &0.0034 &0.0000 &0.0107 &0.0008 &0.0186 &0.0025 &0.0262\\
$6s_{1/2}$&$ 7p_{3/2}$&  0.0471  &0.0082  &0.0232 &0.0000 &0.0137 &0.0027 &0.0087 &0.0084 &0.0058 &0.0150\\
$6p_{1/2}$&$ 7d_{3/2}$&  0.1320  &0.1008  &0.1362 &0.0943 &0.1240 &0.0698 &0.1096 &0.0500 &0.0964 &0.0093\\
$6p_{3/2}$&$ 7d_{5/2}$&  0.1178  &0.1033  &0.1062 &0.0788 &0.0864 &0.0506 &0.0683 &0.0265 &0.0532 &0.0154\\
$6d_{3/2}$&$ 5f_{5/2}$&  1.0723  &0.9891  &1.2019 &1.0900 &1.1694 &0.9986 &0.8777 &0.6713 &0.4842 &0.3307\\
$6d_{5/2}$&$ 5f_{7/2}$&  1.0309  &0.9575  &1.1535 &1.0526 &1.0970 &0.9398 &0.7918 &0.6088 &0.4125 &0.2843\\
$7s_{1/2}$&$ 7p_{1/2}$&  0.4699  &0.4285  &0.4957 &0.4401 &0.4950 &0.4341 &0.4899 &0.4301 &0.4838 &0.4228\\
$6d_{3/2}$&$ 7p_{1/2}$&  0.1058* &0.2544* &0.1118 &0.0946 &0.1921 &0.1797 &0.2363 &0.2245 &0.2635 &0.2504\\
$7s_{1/2}$&$ 7p_{3/2}$&  0.9170  &0.8548  &1.0717 &0.9675 &1.1305 &1.0085 &1.1656 &1.0365 &1.1912 &1.0549\\
$6d_{3/2}$&$ 7p_{3/2}$&  0.0049* &0.0074* &0.0325 &0.0298 &0.0435 &0.0419 &0.0485 &0.0471 &0.0508 &0.0495\\
$6d_{5/2}$&$ 7p_{3/2}$&  0.0340* &0.0814* &0.1890 &0.1719 &0.2617 &0.2515 &0.2970 &0.2878 &0.3157 &0.3068\\
$6d_{5/2}$&$ 5f_{5/2}$&  0.0515  &0.0479  &0.0577 &0.0526 &0.0551 &0.0472 &0.0400 &0.0306 &0.0208 &0.0141\\
$5f_{5/2}$&$ 7d_{3/2}$&  0.0432  &0.0084  &0.1423 &0.1277 &0.1627 &0.1514 &0.1034 &0.0880 &0.0398 &0.0141\\
$7p_{1/2}$&$ 7d_{3/2}$&  0.4801  &0.2558  &0.8913 &0.7706 &1.1480 &1.0501 &1.3123 &1.1758 &1.4230 &0.3889\\
$7p_{3/2}$&$ 7d_{3/2}$&  0.0674  &0.0437  &0.1103 &0.0982 &0.1334 &0.1231 &0.1465 &0.1312 &0.1542 &0.0292\\
$6p_{3/2}$&$ 7d_{3/2}$&  0.0130  &0.0115  &0.0112 &0.0082 &0.0086 &0.0048 &0.0064 &0.0026 &0.0047 &0.0003\\
$5f_{5/2}$&$ 7d_{5/2}$&  0.0036  &0.0013  &0.0107 &0.0096 &0.0115 &0.0096 &0.0069 &0.0056 &0.0024 &0.0023\\
$5f_{7/2}$&$ 7d_{5/2}$&  0.0540  &0.0194  &0.1603 &0.1448 &0.1713 &0.1442 &0.1016 &0.0830 &0.0360 &0.0340\\
$7p_{3/2}$&$ 7d_{5/2}$&  0.5883  &0.3726  &0.9601 &0.8461 &1.1717 &1.0312 &1.2995 &1.1054 &1.3787 &1.2589\\
$7p_{1/2}$&$ 8s_{1/2}$&  0.3050  &0.2619  &0.2851 &0.2671 &0.2723 &0.2648 &0.2655 &0.2619 &0.2595 &0.2590\\
$7p_{3/2}$&$ 8s_{1/2}$&  0.3291  &0.3012  &0.3489 &0.3322 &0.3532 &0.3433 &0.3595 &0.3527 &0.3646 &0.3591\\
$6p_{1/2}$&$ 8s_{1/2}$&  0.0018  &0.0171  &0.0035 &0.0224 &0.0044 &0.0255 &0.0050 &0.0275 &0.0054 &0.0287\\
$6p_{3/2}$&$ 8s_{1/2}$&  0.0108  &0.0142  &0.0217 &0.0218 &0.0265 &0.0260 &0.0298 &0.0286 &0.0321 &0.0303\\
\end{tabular}
\end{ruledtabular}
\end{table*}

\begin{table*}
\caption{\label{tab1-comp} Oscillator strengths evaluated in the
SD approximation ($f^\text{ SD}$)  for transitions in Au~I and
Au-like ions. The data are compared with theoretical
($f^\text{theor}$) and experimental  ($f^\text{expt}$) results
from Ref.~\protect\cite{au-like} and references therein. }
\begin{ruledtabular}
\begin{tabular}{llllllll}
\multicolumn{1}{c}{} &
\multicolumn{1}{c}{$6s$-$6p_{1/2}$ } &
\multicolumn{1}{c}{$6s$-$6p_{3/2}$ } &
\multicolumn{1}{c}{$6p_{1/2}$-$6d_{3/2}$ } &
\multicolumn{1}{c}{$6p_{3/2}$-$6d_{3/2}$ } &
\multicolumn{1}{c}{$6p_{3/2}$-$6d_{5/2}$ } &
\multicolumn{1}{c}{$6p_{1/2}$-$7s$ } &
\multicolumn{1}{c}{$6p_{3/2}$-$7s$ } \\
\hline
\multicolumn{7}{c}{Au~I} \\
$f^\text{ SD}     $&0.188     &0.408    &0.420   &0.053     &0.477    &0.139     &0.167\\
$f^\text{ theor}  $&0.183     &0.418    &0.450   &0.053     &0.524    &0.152     &0.181\\
$f^\text{ expt}   $&0.176     &0.351    &0.42    &          &0.46     &          &     \\
\multicolumn{7}{c}{Hg~II} \\
$f^\text{ SD}     $&0.214     &0.507    &0.702   &0.082     &0.711    &0.153     &0.197\\
$f^\text{ theor}  $&0.200     &0.493    &0.719   &0.075     &0.741    &0.164     &0.206\\
$f^\text{ expt}   $&0.21      &0.52     &0.62    &          &0.59     &0.15      &     \\
\multicolumn{7}{c}{Tl~III} \\
$f^{SD}     $&0.225     &0.559    &0.820   &0.091     &0.708    &0.159     &0.213\\
$f^\text{ theor}  $&0.213     &0.545    &0.844   &0.083     &0.807    &0.164     &0.222\\
$f^\text{ expt}   $&0.30      &0.76     &        &          &         &          &     \\
\multicolumn{7}{c}{Pb~IV} \\
$f^\text{ SD}     $&0.230     &0.594    &0.885   &0.094     &0.680    &0.163     &0.225\\
$f^\text{ theor}  $&0.221     &0.585    &0.919   &0.088     &0.885    &0.169     &0.233\\
$f^\text{ expt}   $&0.23      &0.61     &0.91    &          &0.76     &          &     \\
\multicolumn{7}{c}{Bi~V} \\
$f^\text{ SD}     $&0.232     &0.619    &0.930   &0.097     &0.706    &0.166     &0.234\\
\end{tabular}
\end{ruledtabular}
\end{table*}

\begin{table}
\caption{\label{tab-life} Lifetimes ${\tau}$ in ns
 of the $6p$  levels
 in Au~I, Hg~II, Tl~III, Pb~IV, and Bi~V.  The
lifetime of the upper level is shown. The corresponding
wavelengths $\lambda$ in \AA~  are also given.
 The data are compared with experimental
results from Ref.~\protect\cite{tl2-life} and references therein.}
\begin{ruledtabular}
\begin{tabular}{llllll}
\multicolumn{1}{c}{Lower}                   &
\multicolumn{1}{c}{Upper}                   &
\multicolumn{1}{c}{$\lambda^{(\rm SD)}$}         &
\multicolumn{1}{c}{$\tau^{(\rm SD)}$}            &
\multicolumn{1}{c}{$\lambda^{\text {expt}}$}&
\multicolumn{1}{c}{$\tau^{\text {expt}}$}   \\
\hline
\multicolumn{6}{c}{Au~I, $Z$=79}\\
$6s_{1/2}$&$ 6p_{1/2}$& 2679     &5.72  &2677     &6.2$\pm$0.2   \\
$6s_{1/2}$&$ 6p_{3/2}$& 2414     &4.28  &2429     &4.7$\pm$0.2   \\
\multicolumn{6}{c}{Hg~II, $Z$=80}\\
$6s_{1/2}$&$ 6p_{1/2}$& 1946     &2.65  &1942     &2.91$\pm$0.11 \\
$6s_{1/2}$&$ 6p_{3/2}$& 1651     &1.61  &1650     &1.80$\pm$0.06 \\
\multicolumn{6}{c}{Tl~III, $Z$=81}\\
$6s_{1/2}$&$ 6p_{1/2}$& 1561     &1.63  &1559     &1.95$\pm$0.06 \\
$6s_{1/2}$&$ 6p_{3/2}$& 1268     &0.862 &1266     &1.06$\pm$0.04 \\
\multicolumn{6}{c}{Pb~IV, $Z$=82}\\
$6s_{1/2}$&$ 6p_{1/2}$& 1313.5   &1.12  &1313.1   &1.11$\pm$0.01 \\
$6s_{1/2}$&$ 6p_{3/2}$& 1028.9   &0.534 &1028.6   &0.52$\pm$0.04 \\
\multicolumn{6}{c}{Bi~V, $Z$=83}\\
$6s_{1/2}$&$ 6p_{1/2}$& 1139.1   &0.837 &1139.5   &0.88$\pm$0.10 \\
$6s_{1/2}$&$ 6p_{3/2}$&  864.2   &0.362 & 864.5   &0.301$\pm$0.016\\
\end{tabular}
\end{ruledtabular}
\end{table}

\begin{table}
\caption{\label{tab-hyp} Hyperfine structure parameters, $A$ (in MHz)
for the $6s$ and $6p$ levels in Hg~II and Tl~III.
 The data are compared with theoretical and experimental
results from Ref.~\protect\cite{hg1} - ($a$),
 Ref.~\protect\cite{tl2} - ($b$) and references therein.}
\begin{ruledtabular}
\begin{tabular}{rrrrrrr}
\multicolumn{1}{c}{Level} &
\multicolumn{1}{c}{$A^{(\rm DF)}$} &
\multicolumn{1}{c}{$A^{(3)}$}  &
\multicolumn{1}{c}{$A^{(\rm SD)}$} &
\multicolumn{1}{c}{$A^{(\rm theor)}$} &
\multicolumn{1}{c}{$A^{(\rm expt)}$} \\
\hline
\multicolumn{6}{c}{$^{199}$Hg~II, $I$=1/2, $\mu$=0.5058852 \protect\cite{web}}\\
$6s_{1/2}$&  34072  &  43013  & 41996   &   42366$^a$ &     40460$^a$\\
$6p_{1/2}$&  5536   &  7103   &  7129   &    7116$^a$ &      6870$^a$\\
$6p_{3/2}$&  456    &   702   &  654    &     659$^a$ &          \\
\multicolumn{6}{c}{$^{201}$Hg~II, $I$=3/2, $\mu$=-0.560225 \protect\cite{web}}\\
$6s_{1/2}$&  -12578 &  -15878 &  -15499 &  -15527$^a$ &    -14960$^a$\\
$6p_{1/2}$&  -2044  &  -2622  &  -2631  &   -2608$^a$ &     -2610$^a$\\
$6p_{3/2}$&  -169   &   -259  &  -241   &    -241$^a$ &          \\
\multicolumn{6}{c}{$^{205}$Tl~III, $I$=1/2, $\mu$=1.6382135 \protect\cite{web}}\\
$6s_{1/2}$&  152604 &  182711 &  179852 &   182990$^b$&    181670$^b$\\
$6p_{1/2}$&   30099 &   36496 &   36300 &    36750$^b$&     48870$^b$\\
$6p_{3/2}$&    2532 &    3560 &   3357  &    3339$^b$ &      4800$^b$\\
\end{tabular}
\end{ruledtabular}
\end{table}


\end{document}